\documentclass[aps,prl,twocolumn,groupedaddress,amsmath,amssymb,nofootinbib]{revtex4-2}
\usepackage[utf8]{inputenc}
\usepackage{graphicx,color}
\usepackage{caption}
\usepackage{subcaption}
\usepackage{amsmath,amssymb}
\usepackage{epstopdf}
\usepackage{soul}
\usepackage{tabularx}
\newcommand{\D}{\mathrm{d}}
\newcommand{\e}{\mathrm{e}}
\newcommand{\half}{\frac{1}{2}}

\newcommand{\be}{\begin{equation}}
\newcommand{\ee}{\end{equation}}
\newcommand{\bea}{\begin{eqnarray}}
\newcommand{\eea}{\end{eqnarray}}
\newcommand{\ba} {\begin{align} }
\newcommand{\ea} {\end{align} }

\newcommand{\fc}{f_\mathrm{c}}
\newcommand{\fm}{f_\mathrm{m}}
\newcommand{\rc}{\mathrm{c}}
\newcommand{\m}{\mathrm{m}}
\newcommand{\rl}{\mathrm{l}}
\newcommand{\g}{\mathrm{g}}

\newcommand{\vecr}{\boldsymbol{r}}

\newcommand{\vecp}{\boldsymbol{p}}
\newcommand{\vecn}{\boldsymbol{n}}
\newcommand{\vecj}{\boldsymbol{j}}
\newcommand{\vecv}{\boldsymbol{v}}

\newcommand{\vecx}{\boldsymbol{x}}

\newcommand{\tenQ}{\boldsymbol{Q}}
\newcommand{\tenI}{\boldsymbol{I}}

\newcommand{\tensig}{\boldsymbol{\sigma}}

\newcommand{\ddt}{\partial_t}

\newcommand{\ra}[1]{\textcolor{black}{#1} } 

\begin{document}


\title{\ra{Environment-stored memory in active nematics and extra-cellular matrix remodeling}}

\author{Ram M. Adar$^1$ and Jean-François Joanny$^{2, 3, 4}$}
\email{radar@technion.ac.il}
\affiliation{$1$ Department of Physics,Technion – Israel Institute of Technology, Haifa 32000,Israel \\ 
$^2$ Collège de France, 11 place Marcelin Berthelot, 75005 Paris, France\\
$^3$ Laboratoire Physico-Chimie Curie, Institut Curie, Centre de Recherche, Paris Sciences et Lettres Research University, Centre National de la Recherche Scientifique, 75005 Paris, France \\ 
$^4$ Université Pierre et Marie Curie, Sorbonne Universités, 75248 Paris, France}

\begin{abstract}
Many active systems display nematic order, while interacting with their environment. \ra{In this Letter, we show theoretically how environment-stored memory acts an effective external field that aligns active nematics. The coupling to the environment leads to substantial modifications of the known phase diagram and dynamics of active nematics, including nematic order at arbitrarily low densities and arrested domain coarsening.}
We are motivated mainly by cells that remodel fibers in their extra-cellular matrix (ECM), while being directed by the fibers during migration. Our predictions indicate that remodeling promotes cellular and ECM alignment, and possibly limits the range of ordered ECM domains, in accordance with recent experiments.
\end{abstract}

\maketitle
Active nematics are systems composed of self-propelling constituents \ra{capable of aligning along a shared axis with no preferred overall direction}. The active isotropic-nematic transition has been studied extensively~\cite{Chate2020,Peshkov2012,Ngo2014,Grossmann2016}. Similar to passive liquid crystals, order is driven by strong aligning fields, obtained by a combination of strong interactions and high densities. Unlike passive systems, 
activity couples order with propulsion and allows for coexistence between a dilute isotropic phase and dense nematic phase.  

  Active nematics are ubiquitous in biological systems at different scales. Our main motivation is cells in extra-cellular matrix (ECM), which are both capable of displaying nematic order. Growing biological evidence suggests that the interplay between cellular and ECM order is essential for tissue patterning and multicellular migration~\cite{Clark2015,Alexander2016,VanHelvert2018,Wershof2019,Park2020}. In particular, aligned collagen structures have been shown to greatly promote metastasis~\cite{Conklin2011,Kopanska2016}. 

  \ra{Cell-ECM coupling is especially evident in fibroblasts that deposit, degrade and re-arrange ECM fibers~\cite{Phillips2009,Grinnell2003}. This has been modeled in different contexts, including wound healing~\cite{Olsen1997}, fibroblast alignment~\cite{Li2017}, and ECM patterning~\cite{Dallon1999,Wershof2019}. However, the macroscopic physical mechanisms underlying cell-environment interplay and their role in determining orientational order and dynamics are not well understood or quantified. }
    
Our approach to understand cell-environment interplay is to consider them as a two-component active system. We recently applied such a description to explain {\it mechanical} feedback mechanisms between cells and ECM~\cite{AdarPRL,AdarNJP}. Here we focus on {\it chemical} remodeling. \ra{We find that environment-stored memory acts as an external field that} allows for steady-state nematic order at arbitrarily low densities and constrains angular dynamics. \ra{We relate our results to recent {\it in-vitro} experiments on fibroblasts~\cite{Wershof2019,Park2020}.} While we are motivated by cells in ECM, our \ra{findings are generic and imply that the understanding of standard active matter may not apply in a dynamic environment, highlighting the need for further investigation and adaptation of existing theories}. 

{\it Theory.} We consider active cells and passive environment (matrix) segments in two dimensions, each described by their position and orientation, $\vecr$ and $\vecn$ for the cells and $\vecr'$ and $\vecn'$ for the matrix. Cells self-propel with a velocity $\vecv=v\vecn$ and diffuse with a diffusion coefficient $D$. They also align with neighboring cells and matrix segments. Matrix segments are considered to be apolar. They are enslaved to the cells that may deposit and degrade them \ra{(for more general choices, see SM in~\cite{SM})}.  These dynamics are described by the following equations:
\begin{align}
\label{eq1}
\ddt \fc &=-\nabla\cdot\left(\fc v \vecn\right)+D\nabla^2\fc-k\fc+k\rho_{\rc} \e^{-E_{\rc}}/Z_{\rc}\nonumber\\
\ddt \fm&=\frac{k_+}{2}\left[\fc\left(\vecr',\vecn'\right)+\fc\left(\vecr',-\vecn'\right)\right]-k_-\rho_{\rc} \fm,
\end{align}
 where $\ddt$ denotes the partial time derivative. The function $\fc$ ($\fm$) describes the distribution to find a cell (matrix segment) at position $\vecr$ ($\vecr'$) with orientation $\vecn$ ($\vecn'$). They are normalized such that $\int\D\vecn\fc=\rho_\rc$ is the cellular density and $\int\D\vecn'\fm=\rho_\m$ is the matrix density. 
 
 The cellular orientation dynamics are written in terms of a tumbling rate $k$, and an orientation probability, given by the Boltzmann factor $\exp\left(-E_\rc\right)/Z_\rc$ with the effective alignment energy $E_\rc$ and partition function $Z_\rc=\int\D\vecn\exp\left(-E_\rc\right)$~\cite{Hardrods}.  This is a convenient choice that allows for the recovery of passive systems in simple limits. 
 
 Matrix deposition and degradation are described by the rates $k_+\rho_\rc$ and $k_-\rho_\rc$, respectively. Here we assume that cells locally deposit segments along their axis of motion and degrade segments in all orientations.  \ra{Similar ingredients of cell and matrix dynamics were recently proposed as part of a two-layer Viscek model~\cite{Wershof2019}. We note that Eq.~(\ref{eq1}) is written within mean field.}

Averaging the different moments of the orientation angles yield mesoscopic fields that are the focus of our theory. The active cellular current density is given by  $\vecj=v\int\D\vecn\fc$, the cellular nematic tensor density is $\tenQ_\rc=\int\D\vecn\,\left(\vecn \vecn-\tenI/2\right)\fc$, and the matrix nematic tensor density  is
$\tenQ_\m=\int\D\vecn'\,\left(\vecn' \vecn'-\tenI/2\right)\fm.$  These fields are all extensive in the number of cells/matrix segments.

We coarse-grain Eq.~(\ref{eq1}) into equations in terms of the average fields, \ra{using an approximation that neglects higher moments of $\fc$ in $\vecn$ beyond the nematic tensor}. In particular, we treat the orientation within mean field in terms of the interaction $E_\rc\left(\vecn\right) =-2\text{Tr}\left[\left(\vecn\vecn-\tenI/2\right)\tenQ_t\right]$ with the total aligning field $\tenQ_t=\beta_\rc\tenQ_\rc+\beta_\m\tenQ_\m$. It includes cell-cell and cell-matrix alignment, with the interaction strengths $\beta_\rc$ and $\beta_\m$, respectively (more general choices are given in the SM~\cite{SM}). In the absence of cell activity and cell-matrix interaction, our choice of $E_\rc$ leads to an equivalent of Maier-Saupe theory~\cite{LCdeGennes} for compressible two-dimensional systems.
 
The resulting field equations are~\cite{SM}: 
\begin{align}
\label{eq2}
\ddt \rho_\rc&=D\nabla^2\rho_\rc-\nabla\cdot\vecj,\nonumber\\
\ddt \vecj &=D\nabla^2\vecj-v^2\nabla\rho_\rc/2-v^2\nabla\cdot\tenQ_\rc-k\vecj,\nonumber\\
\ddt \tenQ_\rc &=D\nabla^2\tenQ_\rc-\left(\nabla\vecj+\nabla\vecj^\mathrm{T}-\nabla\cdot\vecj\tenI \right)/4 \nonumber\\
&-k\,\tenQ_\rc +k\rho_\rc \,g\left(Q_t\right) \tenQ_t/Q_t,\nonumber\\
\ddt \rho_\m&=\rho_\rc\left(k_+-k_-\rho_\m\right),\nonumber\\
\ddt\tenQ_\m&=k_+\tenQ_\rc-k_-\rho_\rc \tenQ_\m.
\end{align}
The first equation is the cellular continuity equation, given by the active cellular current $\vecj$ and passive diffusive current. The second equation is a polarization-rate equation for the active current, which we interpret below, at steady state, as a force balance equation. 

The equation for $\tenQ_\rc$ includes diffusion and shear-alignment (first line), as well as non-linear alignment terms that dominate at large lengthscales (second line). They are written in terms of the function $g(x)=I_1(x)/I_0(x),$ where $I_n(x)$ is the modified Bessel function of the first kind~\cite{AS}, \ra{which results from an angular average of the Boltzmann factor $\exp\left(-E_\rc\right)$.}
The cellular dynamics include the first and second moments of the angular distribution ($\vecj$ and $\tenQ_\rc$, respectively), similarly to ``self-propelled rods''~\cite{Marchetti2013,Peruani2006,Baskaran2008}. Finally, the matrix dynamics are governed by cellular deposition and degradation.

\begin{figure}[ht]
\centering
\includegraphics[width=0.85\columnwidth]{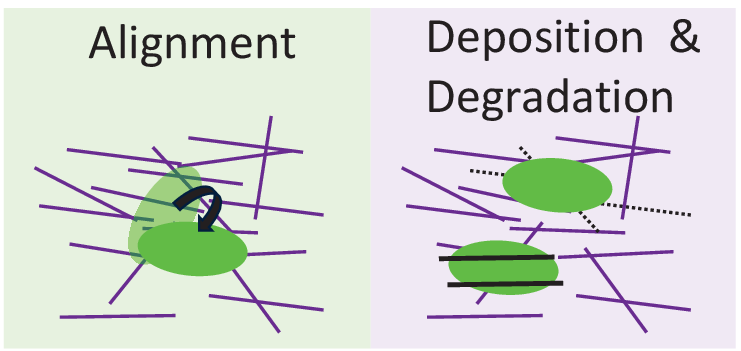}
\caption{(Color online) Heuristic description of cell-matrix feedback. Left panel: Cells (green) align with matrix segments (purple). Right panel: cells degrade exisitng segments (dashed black) and deposit new segments (bold black). The feedback between these processes drives the phenomena in our theory.}
\label{fig1}
\end{figure}

These equations define our framework for active nematics (cells) \ra{with environment-stored memory (matrix nematic order)}, which we apply for the study of ECM remodeling. Cell-matrix interplay enters the theory in two ways: cellular alignment by the matrix as part of the nematic tensor $\tenQ_t$ and  matrix remodeling by the cells (see Fig.~\ref{fig1}). Cellular activity  enters our theory in the active current $\vecj$, matrix deposition and degradation, and possibly in the alignment dynamics.

Next, we focus on the consequences of remodeling on the emergence of cellular and ECM orientational order at steady state as well as typical relaxation dynamics of the cell and matrix. For brevity, we rescale times with the run time $1/k$ and lengths with the typical cellular persistence length $v/k$, while keeping the same notation.

{\it Results.} 
The standard isotropic-nematic transition in active systems is similar to a gas-liquid transition~\cite{Chate2020,Solon2013}, where the alignment strength  plays the role of inverse temperature. At low densities and high temperatures, the system forms a dilute isotropic gas, while at high densities and low temperatures - a nematic liquid. At intermediate densities and temperatures, the two phases co-exist and are generally linearly unstable. Here, we show how the matrix can break this behavior. 

The key to understand the coexistence lies in the stress. In the hydrodynamic limit of large system size and long time, \ra{the total cellular current is proportional to a divergence of a tensor that we interpret as the stress}~\cite{SM}, $\tensig=-\left[\ra{\rho_\rc}\tenI+2\tenQ_\rc/\left(1+2D\right)\right]$.  The steady-state behavior of the cells is thus described by a constant stress tensor. We consider a possible density profile along the $x$ direction and focus on the $xx$ component of the stress that we denote as $\sigma$ for brevity, 
\begin{align}
    \label{eq3}
\sigma=\sigma_{xx}=-\left(\rho_\rc+\frac{Q_\rc}{1+2D}\right).
\end{align}
The first term is the ideal-gas contribution to the pressure, while the second term is an extensile active stress $\sim\tenQ_\rc$~\cite{FibroNote}. Here we consider ordering either along the $x$ axis ($Q_\rc >0$) or the $y$ axis ($Q_\rc <0$). 

Co-existence is possible when the active stress decreases with density,  compensating for the increase in ideal-gas pressure. This is the case for alignment in the $y$ direction. The stress $\sigma$ can be considered as a Lagrange multiplier that enforces the total number of cells. It is given by \ra{(minus)} the density in the isotropic phase. 

\begin{figure*}[ht]
\centering
\includegraphics[width=0.95\textwidth]{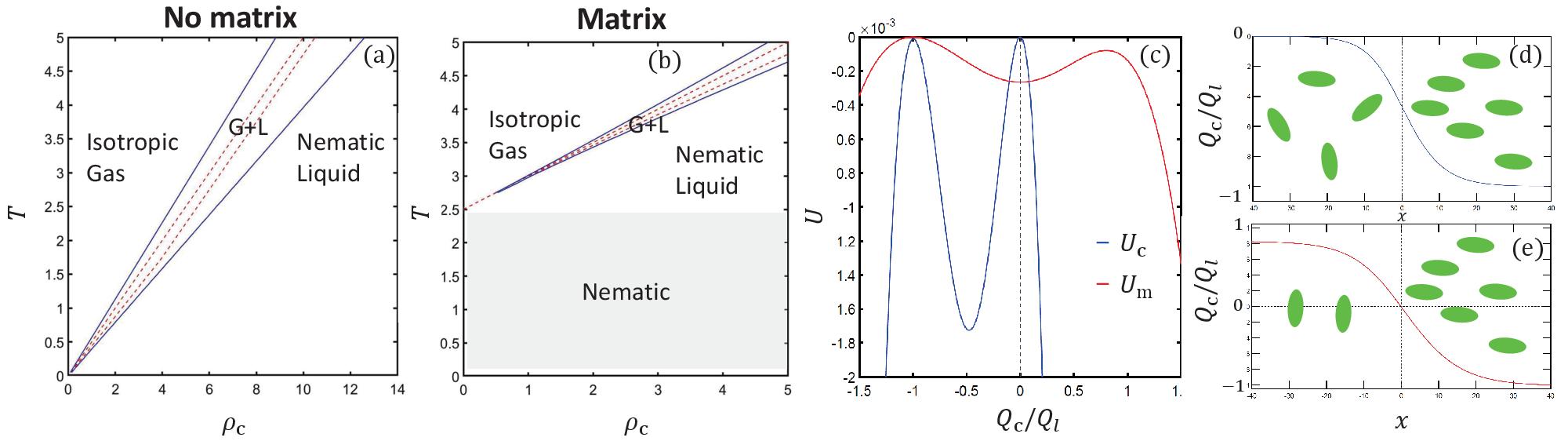}
\caption{(Color online) (a+b) Phase diagrams in the density and temperature plane (a) without a matrix and (b) with a matrix.  We consider $\rho=\rho_\rc$ and $T=1/\beta_\rc$. Solid blue lines are the binodal and dahsed red ones are the spinodal. The values used are: $D=0.5$, $\beta_\m,\rho_\m=0$ (a) and $D=0.5$, $\beta_\m=\beta_\rc,\rho_\m=5$ (b).(c) Comparison between cell-dominated and matrix-dominated potentials. (d+e) Snapshots of co-existence curves from a numerical solution to the hydrodynamic equations [Eq.~(\ref{eq2})] in the cell-dominated (d) and matrix-dominated cases (e). \ra{The green ellipses are a heuristic description of cellular orientational order.} The values used are: $D=0.5$, $\sigma=-1$, and $\beta=2.05$ ($\beta_\rc$ in cell-dominated case and $\tilde{\beta_\m}$ in matrix-dominated case). }
\label{fig2}
\end{figure*}

Next, we derive the isotropic-nematic phase diagram in the density-temperature plane, where $\beta_\rc, \beta_\m\sim1/T$, \ra{and the ratio $\beta_\rc/\beta_\m$ is kept fixed}. Examples of such phase diagrams with and without a matrix (ECM) are given in Figs.~\ref{fig2}a and \ref{fig2}b. The region of co-existence  is delimited by the binodal line (solid blue line), within which lies a region of linear instability, delimited by the spinodal line (dashed red line). 

\ra{{\bf Steady-state nematic order: matrix aligns cells at arbitrarily low densities.}}
We solve Eq.~(\ref{eq2}) at steady state. The matrix density is $\rho_\m = k_+/k_-$, independent of $\rho_\rc$.  The matrix nematic tensor has the same direction as the cellular one, chosen here as the $x$-axis. We define the {\it intensive} nematic order of the cells and matrix, $q_\rc=Q_\rc/\rho_\rc$ and $q_\m=Q_\m/\rho_\m$, and find that $q_\m=q_\rc$ at steady state. 

The matrix thus inherits the same intensive nematic order as the cells. \ra{Consequently $Q_t=\left(\beta_\rc\rho_\rc+\beta_\m\rho_\m\right)q_\rc$ at steady state, and the cellular nematic tensor solves} 
\begin{align}
    \label{eq4}
\ra{q_\rc=g\left[\left(\beta_\rc\rho_\rc+\beta_\m\rho_\m\right)q_\rc\right].}
\end{align}
\ra{This is one of our main results. By expanding the right-hand-side of Eq.~(\ref{eq4}), we find that nematic order is possible for $\beta_\rc\rho_\rc+\beta_\m\rho_\m>2.$ The $\beta_\m\rho_\m$ term quantifies the matrix contribution and allows for nematic order even for vanishing cellular densities $\rho_\rc\approx0$ (grey region in Fig.~\ref{fig2}b). The mechanism is simple: even dilute cells deposit a finite-density matrix after sufficiently long time. The matrix then acts as an external field that aligns the cells. Alternatively, rather than being aligned by current neighbors, cells are aligned by the memory of past neighbors, recorded by the matrix.}

Next, we analyze the effect of ECM remodeling on the spinodal and binodal lines, as is plotted in Fig.~\ref{fig2}.

{\bf Spinodal: \ra{matrix stabilizes the nematic order.}} The spinodal is given by $\partial \sigma/\partial \rho_\rc=0$ for fixed values of $\beta_\rc$ and $\beta_\m$~\cite{SM}. This threshold of linear instability is due to a negative compressibility. As the cellular density increases, the active stress overcomes the osmotic pressure and pushes cells up their concentration gradient. Note that active nematics can also be unstable due to a combination of active stress and shear alignment~\cite{Voituriez2005,Duclos2018}, but this is not the case here, where the cells are effectively extensile and align with the strain rate. 

Negative compressibility occurs for $\partial Q_\rc/\partial \rho_\rc<-\left(1+2D\right)$ [Eq.~(\ref{eq3})]. In the isotropic state, $Q_\rc\equiv0$ and this is not possible. Deep in the ordered state, $Q_\rc=\pm\rho_\rc$, also ensuring stability. The instability is possible, therefore, only for intermediate $Q_\rc$ values. For such values we expand the nonlinear terms of Eq.~(\ref{eq4}) and find its possible roots. One solution is $q_\rc=0$ and the other is $ q_\rc=-\sqrt{\left(\beta_\m\rho_\m+\beta_\rc\rho_\rc-2\right)/\left(\beta_\m\rho_\m+\beta_\rc\rho_\rc\right)^3}.$ 

First, we examine the case of $\beta_\m\rho_\m<2.$ The cells are isotropic at low densities and become ordered at $\rho_\ast=\left(2-\beta_\m\rho_\m\right)/\beta_\rc$. As $Q_\rc\sim\sqrt{\rho-\rho_\ast}$ in this case, $\partial Q_\rc/\partial \rho_\rc\ll-1$ and the system is unstable. The cell density $\rho_\ast$ thus marks the gas spinodal line. Otherwise, for $\beta_\m\rho_\m>2$, the slope $\partial Q_\rc/\partial \rho_\rc$ at vanishing densities is given by $\sqrt{\left(\beta_\m\rho_\m-2\right)/\left(\beta_\m\rho_\m\right)^{3/2}}<1$. The matrix thus increases the compressibility and ensures stability.  \ra{This is why the spinodal lies outside the grey region in Fig.~\ref{fig2}b.}

{\bf Binodal: \ra{matrix allows for co-existence between different orientations.}} The binodal describes, for a given temperature, the densities of the macroscopic phases at coexistence. We find it from the equation for $Q_\rc,$ while replacing $\rho_\rc$ by its  steady-state value, $-\sigma-Q_\rc/\left(1+2D\right)$. Upon proper rescaling of lengths~\cite{SM}, we find that 
\begin{align}
    \label{eq5}    Q_\rc''=Q_\rc+\left(\sigma+\frac{Q_\rc}{1+2D}\right)g\left(Q_t\right)\equiv F\left(\sigma,Q_\rc\right).
\end{align}
This has the same structure as Newton's equation, where $Q_\rc$ plays the role of position and the $x$ coordinate - the role of time, while $F$ is the force (see also~\cite{Solon2015}). The first integral (conservation of energy) yields $E=Q_\rc '^2/2+U$, where we have denoted the \ra{``potential energy''} $U=-\int\D Q_\rc\,F\left(\sigma,Q_\rc\right).$ 
 
\ra{Co-existence requires two $Q_\rc$ values that have the same ``potential energy'' $U$. The co-existing phases can be either finite-sized or macroscopic, depending on the value of $F$. Macroscopic phases occur for} $F=0$, where it takes an infinite ``time'' for the Newtonian particle to switch between the phases. These two conditions set $Q_\rl$, the nematic order in the dense liquid phase, as well as $-\sigma=\rho_{\g}$, the density in the isotropic gas phase. To summarize, we require that $Q_\rc=0, Q_\rl$ are \ra{equally-valued maxima} of $U$ at the binodal. 

 We highlight the effect of the environment by focusing on two limits: a cell-dominated interaction $U\left(\beta_\m=0\right)=U_\rc$ where there is no matrix, and a matrix-dominated one $U\left(\beta_\rc=0\right)=U_\m$, where the cells are aligned only by the matrix. Explicitly, 
\begin{align}
    \label{eq6}
    U_\rc \left(Q_\rc\right)&=\int _0 ^{Q_\rc} \D Q \left[\rho_\rc (Q) g\left(\beta_\rc Q\right)-Q\right],\nonumber\\
U_\m\left(Q_\rc\right)&=\int_0 ^{Q_\rc} \D Q \left[\rho_\rc (Q) g\left(\tilde{\beta_\m}q_\rc(Q)\right)-Q\right],
\end{align}
where $\tilde{\beta_\m}=\rho_{\m} \beta_\m.$ The difference between the two cases is the magnitude of the total nematic tensor ($Q_t$), which appears as the argument of the nonlinear $g$ function. In the cell-dominated case, the argument scales as the {\it extensive} $Q_\rc$ that vanishes at small densities, while the matrix-dominated cases - as the {\it intensive} $q_\rc$. The two potentials are plotted in Fig.~\ref{fig2}c.

\ra{The intensive nematic order $q_\rc$ in the cell-dominated case is a function of $\beta_\rc\rho_\rc$ [Eq.~(\ref{eq4})]} and both the spinodal and binodal lines are given by $\beta_\rc \rho_\rc=\mathrm{const,}$ as is displayed on Fig.~\ref{fig2}a. In particular, we find that the nematic order at the liquid binodal $\beta_\rc Q_\rl$ is not necessarily small~\cite{SM}. Therefore, we cannot find it from an expansion of $U_\rc$, but rather from its full nonlinear form \ra{that we evaluate numerically (and see Fig.~\ref{fig2}c)}. We find that there is indeed a macroscopic coexistence between an isotropic gas and nematic liquid, obtained from the maxima of 
 $U_\rc$ \ra{for a specific value of $\rho_\g$.} The value $\rho_\rl$ is then found by requiring a fixed stress, {\it i.e.}, $\rho_\g=\rho_\rl+Q_\rl/\left(1+2D\right)$.  \ra{Co-existence was validated by numerical solutions of Eq.~(\ref{eq2}) in 1D~\cite{NumericsNote}}, plotted in Fig.~\ref{fig2}d. 

The situation is very different in the matrix-dominated case. The value of $q_\rc$ in this case depends only on $\tilde{\beta_\m}$ [Eq.~(\ref{eq4})]. We expand for small $Q_\rc$ and find $U_\m\sim-Q_\rc^2\left[Q_\rc^2-16\sigma^2\tilde{\beta_\m}^{-3}\left(-2+\tilde{\beta_\m}\right)\right]$. In this case, $Q_\rc=0$ is a local minimum and the global maxima are $Q_\rc=\pm2\sigma\sqrt{\tilde{\beta_\m}^{-3}\left(-2+\tilde{\beta_\m}\right)}$. 

\ra{Equation~(\ref{eq4}) ensures that for any solution $q_\rc=q$ of $F=0,$ $q_\rc=-q$ is also a solution. It can be shown analytically~\cite{SM} that $q_\rc<0$ is the global maximum, while $q_\rc>0$ is a local one,  as is demonstrated by a numerical plot of $U_\m$ in Fig.~\ref{fig2}c. This form of $U_\m$ allows for co-existence between finite domains with nematic order in the $x$ and $y$ directions.} \ra{For example, a nematic order $q_\rc=q>0$, forced by surface anchoring, will transition to $q_\rc=-q$ in the bulk, along a thickness that diverges logarithmically with $\tilde{\beta_\m}-2$~\cite{SM}.}

\ra{The co-existence between differently-oriented domains is verified by numerical solutions of Eq.~(\ref{eq2}) in 1D~\cite{NumericsNote}, plotted in Fig.~\ref{fig2}e.} This new type of coexistence is possible because cells order at arbitrarily low densities. Then, cells aligned along the $x$ direction at very low densities can exert a positive active stress that matches $\sigma$. \ra{The exact form of co-existence profiles depends on angular dynamics, as explained next.}

{\bf Angular dynamics: \ra{matrix possibly arrests domain coarsening.}}
Finally, we focus on angular dynamics. While the system is invariant under global rotations of the cells and matrix together, their preferred mutual alignment results in a finite relaxation rate of their relative angle that is independent of system size. We define the angle between the preferred axis of the cells and the $x$ axis as $\phi_\rc$ \ra{such that the two independent terms in $\tenQ_\rc$ are $Q_\rc\cos (2\phi_\rc)/2$ and $Q_\rc\sin (2\phi_\rc)/2$. We similarly define $\phi_\m$ for the matrix}. The relative angle between them is $\alpha/2=\phi_\rc-\phi_\m$. \ra{We rewrite Eq.~(\ref{eq2}) in terms of $Q_\rc$, $Q_\m$, $\phi_\rc$ and $\phi_\m$, and find that}~\cite{SM}. 
\begin{align}
   \label{eq7}
\partial_{t}\alpha & =-\left[k\beta_{\m}\rho_\m\frac{q_{\m}}{q_{\rc}}\frac{g\left(Q_{t}\right)}{Q_{t}}+k_{+}\frac{\rho_\rc}{\rho_\m}\frac{q_{\rc}}{q_{\m}}\right]\sin\alpha,
\end{align}
 where we have included the time scale $1/k$ explicitly. Note that all the densities and nematic orders also evolve in time and are coupled with $\alpha$, {\it e.g.}, via shear alignment.

The two terms in the parenthesis on the right-hand side of Eq.~(\ref{eq7}) describe the dynamics of the cells and matrix, respectively. In the ordered state, their characteristic rates scale as $k \beta_\m  \rho_\m/\left(\beta_\m  \rho_\m+\beta_\rc  \rho_\rc\right)$ and $k_-\rho_\rc,$ respectively~\cite{SM}. The cellular rate depends on the typical cellular re-orientation rate and the strength of its alignment to the matrix field, while the matrix rate is defined by the degradation rate. The interplay between these two rates determines whether the cells are free to rotate with the matrix constantly remodeling according to the cells $[k_-\rho_\rc\gg k \beta_\m  \rho_\m/\left(\beta_\m  \rho_\m+\beta_\rc  \rho_\rc\right)]$ or the cells are pinned to the matrix $[k_-\rho_\rc\ll k \beta_\m  \rho_\m/\left(\beta_\m  \rho_\m+\beta_\rc  \rho_\rc\right)]$. 

The latter implies that suppression of cellular relaxation dynamics. For example, consider ordered cellular domains of typical size $l$ with different orientations \ra{(different $\phi_\rc$ values), such as alternating bands of width $l$}. As long as $k \beta_\m  \rho_\m/\left(\beta_\m  \rho_\m+\beta_\rc  \rho_\rc\right)\gg D_t/l^2,\,k_-\rho_\rc$, we expect these domains to remain frozen, rather than relax into a common orientation, as is the usual case \ra{(see Supplemental Figure in~\cite{SM})}. Here we have denoted $D_t$ as the total translational diffusion coefficient. In our model, it is given by $D_t=D+v^2/(4k)$.


{\it Discussion.}
\ra{This work demonstrates how environment-stored memory qualitatively changes the known behavior of active nematics. The underlying mechanism is generic: active particles generate a finite external field even for vanishing densities.  Our findings open an avenue for novel behavior of active systems. Arrested domain coarsening, for example, suggests that the steady state may contain a signature of the initial conditions}. Environment-induced relaxation dynamics should also slow down defect dynamics (as was shown very recently in ~\cite{Jacques2023}) and possibly arrest typical instabilities, such as nematic bands at coexistence~\cite{Chate2020} \ra{and flow transitions~\cite{Voituriez2005}. Finally, this may also decrease the role of fluctuations beyond mean field}. 

Our finding are useful in understanding ECM remodeling by cells and its consequences on  cellular and tissue dynamics. \ra{We focus on  quasi-2D, {\it in-vitro} studies of fibroblasts and their derived matrices (see, {\it e.g.}, Ref.~\cite{Park2020}.) 
The cells exchange momentum with the underlying substrate, as is the case in ``dry'' active systems. The rigid substrate also suppresses elastic matrix deformations. Nevertheless, ECM displays orientational order for cellular densities of the order $10^{-4}\mu\mathrm{m}^{-2}$, which correspond to $\rho_\rc\approx10^{-2}$, as can be understood from the memory effect in our theory (see also~\cite{Li2017}).} In setups where elasticity is important, it is expected to serve as another mechanism for alignment~\cite{De2007,Livne2014}. \ra{Generally, ECM rheology is complex, including visco-plastic contributions~\cite{Artola2021}.}

It was recently reported~\cite{Wershof2019} that fibroblast-ECM interaction promotes alignment in non-aligned ECMs, but may also decrease the range of alignment. This is explained by our theory in a simple way: increasing the interaction means a larger cellular aligning field $Q_t$,  leading to alignment. At the same time, increasing $\beta_\m$ also increases the rate of cellular relaxation to the matrix, and may thus suppress domain coarsening. \ra{For dilute cells and assuming that the translational diffusion is mainly active ($D_t\sim v^2/k$), we predict a domain size of the order of the cellular persistence length, {\it i.e.,} of the order $10\mu$m. This is consistent with experimental findings~\cite{Park2020,Wershof2019}.}  In a future work, we will further apply our framework to predict ECM patterns observed in vivo.  

\ra{In conclusion, our work demonstrates the profound effect of environment-stored memory on the steady-state and dyanmics of active nematics, especially in the biological context of ECM remodeling. It is generic in nature and is expected to play a similar role in additional active  systems, including polar and synthetic.}

{\it Acknowledgements.} We thank Erik Sahai and Rapha\"{e}l Voituriez for useful discussions.

\bibliography{Refs}

\begin{thebibliography}{37}
\bibitem [{Har()}]{Hardrods}%
  \BibitemOpen
\bibfield  {journal} {  }\href@noop {} {}\bibinfo {note} {As cells are
  deformable and interact among themselves and with ECM via adhesion bonds, we
  describe the alignment via an interaction, $E_\rc$. Generally, alignment can
  also be driven by excluded volume between hard rods.}\BibitemShut {Stop}%
\bibitem [{Fib()}]{FibroNote}%
  \BibitemOpen
  \href@noop {} {}\bibinfo {note} {While our theory corresponds to an extensile
  active stress, cells such as fibroblasts are often contractile. The same
  qualitative results apply in the contractile case. The main difference is
  that the nematic order will form along the $x$ direction, rather than the $y$
  direction.}\BibitemShut {Stop}%
\end{thebibliography}
\onecolumngrid
\section*{Supplemental material for ``Environment-stored memory in active nematics and extra-cellular matrix remodeling''}
This Supplemental Material (SM) provides, in greater detail, the derivation
of the hydrodynamic equations and phase diagram. The outline of the SM is as follows. In
Sec.~\ref{SMsec1}, we coarse grain the microscopic dynamics [Eq. (1) of the Letter] into hydrodynamic equations [Eq. (2) of the Letter]. In Sec~\ref{SMsec2}, we solve the equations at steady-state, analyze the linear stability of the steady-state (spinodal) and derive the conditions for phase coexistence (binodal). Finally, 
in Sec.~\ref{SMsec4}, we derive the nonlinear equations for the angular dynamics.

\section{Derivation of hydrodynamic equations}
\label{SMsec1}
As our starting point, we consider the following microscopic dynamics of the cells and matrix segments [Eq. (1) of the Letter],
\begin{align}
\label{eqs1}
\frac{\partial \fc}{\partial t} &=-\nabla\cdot\left(\fc v \vecn\right)+D\nabla^2\fc-k\fc+k\rho_{\rc} \frac{\e^{-E_{\rc}}}{Z_{\rc}}\nonumber\\
\frac{\partial \fm}{\partial t} &=\frac{k_+}{2}\left[\fc\left(\vecr,\vecn\right)+\fc\left(\vecr,-\vecn\right)\right]-k_-\rho_{\rc} \fm\nonumber\\
&-k_0\fm+k_0\rho_{\m} \frac{\e^{-E_{\m}}}{Z_\m}.
\end{align}
The function $\fc$ ($\fm$) describe the distribution to find a cell (matrix segment) at position $\vecr$ with orientation $\vecn$ ($\vecr'$ with $\vecn'$).
The microscopic equations describe how it changes due to diffusion, advection, and re-orientation. 
\ra{Here, in addition to the terms in Eq.~(1) of the main text, we account for possible re-arrangment of matrix segments by the cells. This is descirbed similarly to cellular alignment, with a typical rate $k_0$ and a Boltzmann factor, defined by the effective matrix interaction energy $E_\m$ and partition function $Z_\m=\int D\vecn'\exp\left(-E_\m\right)$.}
Generally, the rates may depend on the matrix and cellular densities.

Averaging the different moments of the orientation angles yield mesoscopic fields that are the focus of the hydrodynamic equations:
\begin{align}
\label{eqs2}
\rho_\rc\left(\vecr\right)&=\int\D\vecn\,\fc\left(\vecr,\vecn\right),\nonumber \\
\vecj\left(\vecr\right)&=\int\D\vecn\,v\vecn\fc\left(\vecr,\vecn\right),\nonumber \\
\tenQ_\rc\left(\vecr\right)&=\int\D\vecn\,\left(\vecn \vecn-\tenI/d\right)\fc\left(\vecr,\vecn\right),\nonumber \\
\rho_\m\left(\vecr'\right)&=\int\D\vecn'\,\fm\left(\vecr',\vecn'\right),\nonumber \\
\tenQ_\m\left(\vecr'\right)&=\int\D\vecn'\,\left(\vecn' \vecn'-\tenI/d\right)\fm\left(\vecr',\vecn'\right),\nonumber \\
\end{align}
where $d$ is the space dimension ($d=2$ hereafter). We emphasize that these fields are extensive in the number of cells/matrix segments. In particular, the $\tenQ$ tensors are proportional to the density, as compared to the standard definition of nematic tensors. We define also the intensive nematic order parameter of the cells $q_\rc=Q_\rc/\rho_\rc$ and of the matrix $q_\m=Q_\m/\rho_\m$. Note that the first moment of the matrix angle (polarization) vanishes due to our choice of nematic interaction (see next) and absence of matrix advection.

According to Eq.~(\ref{eqs1}), the cells and matrix segments reorient due to effective interactions $E_\rc$ and $E_\m$, respectively. This is a convenient choice that allows for the recovery of passive systems in simple limits. We focus on nematic interactions, which prefer a certain axis but not a direction. Furthermore, we treat the interactions within a mean-field (MF) approximation, such that the energies can be written in terms of the average fields introduced in Eq.~(\ref{eqs2}) as 
\begin{align}
\label{eqs3}
E_\rc\left(\vecn\right) & =-2\text{Tr}\left[\left(\vecn\vecn-\tenI/2\right)\left(\beta_{\rc\rc} \tenQ_\rc+\beta_{\m\rc}\tenQ_\m\right)\right],\nonumber \\
E_\m\left(\vecn'\right)& =-2\text{Tr}\left[\left(\vecn'\vecn'-\tenI/2\right)\left(\beta_{\m\m}\tenQ_\m+\beta_{\rc\m}\tenQ_\rc\right)\right.
\end{align}
Here the coefficients $\beta_{\rc\rc}$ and $\beta_{\m\m}$ describe the strength of the cell-cell and matrix-matrix aligning interactions, respectively, while $\beta_{\m\rc}$ and $\beta_{\rc\m}$ describe how the cells are aligned by the matrix and how the matrix is aligned by the cells, respectively. The two are not necessarily equal in active, non-equilibrium systems (non-reciprocal interaction~\cite{You2020,fruchart2021}). 

Hereafter, we focus on the reciprocal case, such that $\beta_{\m\rc}=\beta_{\rc\m}$. Furthermore, we assume a preferred parallel alignment and that the matrix is enslaved to the cells. We simplify the notations and write
\begin{align}
\label{eqs4}
E_\rc\left(\vecn\right) & =-2\text{Tr}\left[\left(\vecn\vecn-\tenI/2\right)\tenQ_t\right],\nonumber \\
E_\m\left(\vecn'\right)& =-2\text{Tr}\left[\left(\vecn'\vecn'-\tenI/2\right)\beta_\m\tenQ_\rc\right],
\end{align}
where we have defined the "total" nematic tensor that aligns the cells $\tenQ_t=\beta_\rc\tenQ_\rc+\beta_\m\tenQ_\m$. 

Next, we describe the coarse-graining procedure. It is similarly possible in the more general case of Eq.~(\ref{eqs3}) and $d=3$. We reserve this calculation for a future work.

Multiplying Eq.~(\ref{eqs1}) by the appropriate powers of $\vecn$ and $\vecn'$ and carrying out the integration leads to 
\begin{align}
    \label{eqs5}
    \frac{\partial\rho_\rc}{\partial t}&=-\nabla\cdot\vecj+D\nabla^2\rho_\rc,\nonumber\\
    \frac{\partial\vecj}{\partial t}&=-v^2\nabla\cdot\tenQ_\rc-\frac{v^2}{2}\nabla\rho_\rc+D\nabla^2\vecj-k\vecj,\nonumber\\
    \frac{\partial\tenQ_\rc}{\partial t}&=-v\nabla\cdot\langle\vecn\,\left(\vecn\vecn-\tenI/2\right)\rangle+D\nabla^2\tenQ_\rc-k\tenQ_\rc+k\rho_\rc\,g\left(Q_t\right)\frac{\tenQ_t}{Q_t},\nonumber\\
    \frac{\partial\rho_\m}{\partial t}&=\rho_\rc\left(k_+-k_-\rho_\m\right),\nonumber\\
    \frac{\partial \tenQ_\m}{\partial t} &=k_+\tenQ_\rc-k_-\rho_\rc \tenQ_\m-k_0\tenQ_\m +k_0 \rho_\m\,g\left(\beta_\m Q_\rc \right)\frac{\tenQ_\rc}{Q_\rc},
    \end{align}
    where $\left<
    ...\right>$ denotes and angular average with the probability density $\fc$. Here we have defined the function $g(x)=I_1(x)/I_0(x),$ where $I_n(x)$ is the modified Bessel function of the $n$th kind~\cite{AS}.

    The nonlinear terms were obtained by carrying out integrals of the form
    \begin{align}
        \label{eqs6}
        \frac{\int\D\vecn\,\e^{2\text{Tr}\left(\vecn\vecn\cdot\tenQ\right)}\left(\vecn\vecn-\tenI/2\right)}{\int\D\vecn\,\e^{2\text{Tr}\left(\vecn\vecn\cdot\tenQ\right)}}&=\frac{\int _0 ^{2\pi }\D\theta\e^{Q\cos2\theta}\half\left(\begin{matrix}\cos2\theta & \sin 2\theta\\\sin2\theta & -\cos2\theta\end{matrix}\right)}{\int _0 ^{2\pi} \D\theta\e^{Q\cos2\theta}}\nonumber\\
        &=\frac{I_1\left(Q\right)}{I_0\left(Q\right)}\frac{\tenQ}{Q},
    \end{align}
where $\tenQ=\half Q\left(\begin{matrix}1&0\\0&-1\end{matrix}\right)$ is a generic nematic tensor ( $\tenQ=\tenQ_t$ for the cell dynamics and $\tenQ=\beta_\m\tenQ_\rc$ for the matrix dynamics).

The only term for which we do not have an exact expression is the advective term for $\tenQ_\rc$, $\sim\langle\vecn\vecn\vecn\rangle$. It is given by a higher moment of $\vecn$, whose dynamics are determined by an even higher moment and so forth. We close our equations
by considering moments only up to second order  and
by inserting the ansatz
\begin{align}
\label{eqs7}
f_\rc & =f_{0}\rho+f_{1}\vecj\cdot\boldsymbol{n}+f_{2}\text{Tr}\left(\tenQ_\rc\cdot \vecn\vecn\right).
\end{align}
Inserting this expression in Eq.~(\ref{eqs1}) and enforcing the equalities leads to the final expression
\begin{align}
\label{eqs8}
f_\rc & =\frac{\rho}{2\pi}+\frac{1}{\pi v}\vecj\cdot\vecn+\frac{2}{\pi}\text{Tr}\left(\tenQ_\rc\cdot \vecn\vecn\right) .
\end{align}

The above form of the probability density function allows for the calculation of the average
\begin{align}
    \label{eqs9}
  v\nabla\cdot\langle \vecn \left(\vecn\vecn-\frac{\tenI}{2}\right)\rangle&=\frac{1}{2\pi}\int_0 ^{2\pi}\D\theta\,\left(\begin{matrix}
      \cos2\theta&\sin2\theta\\\sin2\theta&-\cos2\theta
  \end{matrix}\right)\left(\cos\theta\partial_x+\sin\theta\partial_y\right)\left(\cos\theta j_x+\sin\theta j_y\right)\nonumber\\
  &=\frac{1}{4\pi}\int_0 ^{2\pi}\D\theta\,\left(\begin{matrix}
      \cos^2 2\theta\left(\partial_xj_x-\partial_y j_y\right)&\sin^2 2\theta\left(\partial_xj_y+\partial_yj_x\right)\\\sin^2 2\theta\left(\partial_xj_y+\partial_yj_x\right)&-\cos^2 2\theta\left(\partial_xj_x-\partial_y j_y\right)
  \end{matrix}\right)\nonumber\\
  &=\frac{1}{4}\left(\nabla\vecj+\nabla\vecj^\mathrm{T}-\nabla\cdot\vecj\tenI\right).
\end{align}
Substituting this result in Eq.~(\ref{eqs5}) yields Eq. (2) of the main text.

Finally, we rewrite the hydrodynamic equations in dimensionless form. For the sake of brevity, we retain the same notations as before. Times are scaled with the average time between cellular tumbles $1/k$, and lengths are scaled with the cellular persistence length, $l=v/k$, the typical distance a cell covers between tumbles. We find that
\begin{align}
    \label{eqs10}
    \frac{\partial\rho_\rc}{\partial t}&=D\nabla^2\rho_\rc-\nabla\cdot\vecj,\nonumber\\
    \frac{\partial\vecj}{\partial t}&=D\nabla^2\vecj-\nabla\cdot\tenQ_\rc-\half\nabla\rho_\rc-\vecj,\nonumber\\
    \frac{\partial\tenQ_\rc}{\partial t}&=D\nabla^2\tenQ_\rc-\frac{1}{4}\left(\nabla\vecj+\nabla\vecj^\mathrm{T}-\nabla\cdot\vecj\tenI\right)-\tenQ_\rc+\rho_\rc\,g\left(Q_t\right)\frac{\tenQ_t}{Q_t},\nonumber\\
    \frac{\partial\rho_\m}{\partial t}&=\rho_\rc\left(k_+-k_-\rho_\m\right),\nonumber\\
    \frac{\partial \tenQ_\m}{\partial t} &=k_+\tenQ_\rc-k_-\rho_\rc \tenQ_\m-k_0 \tenQ_\m +k_0 \rho_\m\,g \left(\beta_\m Q_\rc \right)\frac{\tenQ_\rc}{Q_\rc}.
    \end{align}
        \section{Steady-state phase diagram}
        \label{SMsec2}
The phase diagram in the standard active nematic case (no matrix) is analyzed similarly to a liquid-gas transition in the density-temperature plane, where the alignment strength plays the role of inverse temperature~\cite{Chate2020}. The system forms an isotropic "gaseous" phase at low densities and high temperatures, and a nematic "liquid" phase at high densities at low temperatures. In between, there is a co-existence region that is generally unstable. Here, we explain how the matrix modifies this picture and derive the new phase diagram from Eq.~(\ref{eqs10}).

\subsection{Steady state}
We solve the hydrodynamic equations [Eq.~(\ref{eqs10})] at steady-state. First, we focus on the active cellular current density. For large lengthscales, we neglect the diffusion term and retain 
\begin{align}
    \label{eqs12}
    \vecj=-\half\nabla\cdot\left(\rho_\rc\tenI+2\tenQ_\rc\right).
\end{align}
Substituting in the cellular continuity equation, we find
\begin{align}
    \label{eqs12}
    \frac{\partial\rho_\rc}{\partial t}&=-\left(D+\half\right)\nabla\cdot\left[\nabla\cdot\left(\rho_\rc\tenI+\frac{2\tenQ_\rc}{1+2D}\right)\right].
\end{align}

The right-hand side of this equation is the divergence of the total cellular current density. The expression for it can be interpreted as force balance. The friction due the current is balanced by the divergence of the stress tensor, $\tensig=-\rho_\rc\tenI-2\tenQ_\rc/\left(1+2D\right)$. It includes an ideal-gas-like term and an extensile active stress $\sim\tenQ_\rc$. Within this picture, the cells can be considered as ``pushers''. Similar results can be obtained for "pullers" (contractile active stress), by rotating the nematic tensor by $\pi/2$ (setting $\tenQ_\rc\to-\tenQ_\rc$).

At steady-state, the effective stress tensor is fixed, introducing a constraint between the density and nematic order parameter. We consider variations of these fields along the $x$ axis. The above constraint reduces to 
\begin{align}
    \label{eqs13}
\rho_\rc+\frac{Q_\rc}{1+2D}=-\sigma_{xx}=-\sigma,
\end{align}
 where we have denoted $\sigma_{xx}=\sigma$ for brevity. For now, we consider a homogeneous steady-state, where the stress is fixed by definition. Below, we analyze also possible profiles along the $x$ direction to find the binodal of the phase diagram.

In the homogeneous case, the cellular density is determined by the initial condition. Cellular nematic order may form in any direction and we define its direction as the $x$ direction. Its magnitude is determined by the nonlinear terms in Eq.~(\ref{eqs10}). In order to find it, we  must first determine the steady state of the matrix.

The matrix density is given by $\rho_\m=k_+/k_-$. Importantly, this expression is independent of $\rho_\rc$. Even rather dilute cells can deposit a finite-density matrix after sufficient time. 
The matrix nematic order $\tenQ_\m$ is enslaved to the cellular one and effectively renormalizes cell-cell interactions. It forms along the same axis as the cellular tensor $\tenQ_\rc$. It is given by 
\begin{align}
    \label{eqs18}
    \tenQ_\m &=\frac{\rho_\m}{k_-\rho_\rc+k_0}\left[k_-+k_0\frac{g\left(\beta_\m Q_\rc \right)}{Q_\rc}\right]\tenQ_\rc.
\end{align}

The magnitude $Q_\m$ is proportional to the matrix density, as expected. The first term in the parenthesis of Eq.~(\ref{eqs18}) originates from active matrix deposition \ra{and degradation (note that $\rho_\m$ itself includes $k_+$)}, while the second term - from re-arrangement due to alignment. The weighted contribution of each term is determined by the rates $k_-$ and $k_0$, respectively. We focus hereafter on matrix deposition and set $k_0=0,$ \ra{restricting ourselves to the limit presented in the main text.} In this case, $q_\m= q_\rc$. In any case, the matrix dominates $Q_t$ at low cellular densities.

The magnitude of the cellular nematic order $Q_\rc$ is determined by 
\begin{align}
   \label{eqs21}
   F\equiv Q_\rc-\rho_\rc g\left(Q_t\right)=0,
\end{align} 
While $Q_\rc=0$ is always a solution, a non-vanishing solution also appears for $F_Q(0)\le0,$ where $F_Q=\partial F/\partial Q_\rc$. We similarly define the partial derivative with respect to the density as $F_\rho=\partial F/\partial \rho_\rc$.

We expand $F$ for small $q_\rc$ and find that
\begin{align}
    \label{eqs24}
    F\approx\rho_\rc \left[q_\rc\left(1-\frac{\beta_\rc\rho_\rc+\beta_\m\rho_\m}{2}\right)+\frac{q_\rc^3}{16}\left(\beta_\rc\rho_\rc+\beta_\m\rho_\m\right)^3\right].
\end{align}
 The roots of $F$ are either $Q_\rc=0$ or
 \begin{align}
    \label{eqs24}
    Q_\rc=\pm\rho_\rc\sqrt{\frac{\beta_\m\rho_\m+\beta_\rc\rho_\rc-2}{\left(\beta_\m\rho_\m+\beta_\rc\rho_\rc\right)^3}}.
\end{align}
This means that $q_\rc$ is only a function of the mean field $\beta_\m\rho_\m+\beta_\rc\rho_\rc$.

First, we examine the case $\beta_\m\rho_\m<2.$ The cells are isotropic at low densities and become ordered at $\rho_\ast=\left(2-\beta_\m\rho_\m\right)/\beta_\rc$. We see that $Q_\rc\sim\sqrt{\rho-\rho_\ast}$ in this case. Otherwise, for $\beta_\m\rho_\m>2,$ the cells can be ordered even at vanishing densities. This is the main effect of the matrix: the cells are aligned by the cell-matrix interaction, because the matrix has a finite density and it acts as an external field even at vanishing cellular densities. 
 
\subsection{Linear stability analysis: spinodal}
Next, we analyze the linear stability of the steady-state solution in the limit of infinite wavelength. The onset of instability defines the spinodal line. In this hydrodynamic limit, we can treat the cellular current, as well as the density and nematic order of the matrix as fast variables. The stability analysis, therefore, is restricted to the cellular density and nematic tensor. It is convenient to write the nematic tensor in terms of its magnitude $Q_\rc$ and angle $\phi_\rc$
\begin{align}
    \label{eqs19}
    \tenQ_\rc=\frac{Q_\rc}{2}\left(\begin{matrix}\cos2\phi_\rc&\sin2\phi_\rc\\\sin2\phi_\rc&-\cos2\phi_\rc\end{matrix}\right).
\end{align}

We analyze the linear stability of the steady state with respect to perturbations with a growth rate $s$ and wave vector $\vecp=p\left(\cos\theta,\sin\theta\right)$  of the form $\vecx = \vecx^0 +\vecx^1 \exp\left(st+i\vecp\cdot\vecr\right),$ where
$\vecx = (\rho_\rc,Q_\rc,\phi_\rc)$. 

Linearizing Eq.~(\ref{eqs10}) yields 
\begin{align}
    \label{eqs20}
s\rho_\rc ^1&=-p^2\left[D+\half+\half\cos2\theta Q_\rc^1+\sin2\theta Q_\rc^0\phi_\rc ^1\right],\nonumber\\
s Q_\rc^1&=-p^2\left[\left(D+\frac{1}{4}\right)Q_\rc ^1+\frac{1}{4}\cos2\theta\rho_\rc ^1\right]-F_\rho \rho_\rc ^1-F_Q Q_\rc ^1,\nonumber\\
2Q_\rc ^0 s\phi_\rc ^1&=-p^2\left[\left(D+\frac{1}{4}\right)2Q_\rc ^0\phi_\rc ^1+\frac{1}{4}\sin2\theta\rho_\rc ^1\right],
\end{align} 
where, as before, $F_Q=\partial F/\partial Q_\rc$ and $F_\rho=\partial F/\partial \rho_\rc$.

The equation on $\rho_\rc ^1$ infers that the scaling of the growth rate is $O\left(p^2\right)$. In the hydrodynamic limit $p\ll1$ we retain only such terms of order $p^2$. This means that the contribution of rotations ($\phi_\rc ^1$) to density changes is negligible and that we can write $Q_\rc ^1\approx-F_\rho \rho_\rc^1/F_Q$. Inserting back in the equation for the density and focusing on $\theta=0$, where the destabilizing term is largest, leads to
\begin{align}
    \label{eqs22}
s&=-p^2\left(D+\half-\half\frac{F_\rho}{F_Q}\right).
\end{align}
The system is thus linearly unstable for $\frac{F_\rho}{F_Q}\geq 1+2D,$ where the equality defines the spinodal. 

The interpretation of this instability becomes straightforward when we notice that $-F_\rho/F_Q=-\partial Q_\rc/\partial \rho_\rc$. Eq.~(\ref{eqs13}) then yields
\begin{align}
    \label{eqs23}
\frac{\partial\sigma}{\partial\rho_\rc}\geq 0.
\end{align}
This is, therefore, a mechanical instability. It occurs when the cells become sufficiently ordered upon a density increase, such that the active stress overcomes the pressure and pushes cells up their density gradient. 

\subsection{Analysis of the spinodal criterion}

The instability criterion is by $-\partial Q_\rc/\partial\rho_\rc\geq1+2D$. It can be understood from the functional dependence of $Q_\rc\left(\rho_\rc\right)$. In the absence of a matrix, at a given temperature, the cells are isotropic up to a finite density $\rho_\ast$. Around this density, the nematic order scales as $Q_\rc\sim\sqrt{\rho_\rc-\rho_\ast}$. Therefore, the derivative diverges at this point and the criterion for instability is fulfilled for $Q_\rc<0.$ This defines the gas spinodal. At large densities, all the cells are ordered such that $Q_\rc=-\rho_\rc$ and the system is stable. The density where stability sets in defines the liquid spinodal. 

The matrix may break this behavior for sufficiently strong interactions. As we have found in Eq.~(\ref{eqs24}), the matrix allows for the cells to be aligned at zero density for $\beta_\m\rho_\m>2$. In this case $\partial Q_\rc/\partial \rho_\rc$ is sufficiently small, such that the system is always stable. Otherwise, for $\beta_\m\rho_\m<2$, the gas spinodal is given by $\beta_\rc\rho_\ast=2-\beta_\m\rho_\m$. 

\subsection{Coexistence criteria: binodal}
Consider an isotropic dilute (gas) phase with density $\rho_\g$ and an ordered dense (liquid) phase with density $\rho_\rl$ and nematic order parameter $Q_\rl$. Co-existence requires an equal stress $\sigma$ across the system. This sets
 \begin{align}
 \label{eqs14}
 \rho_\g&=-\sigma=\rho_{\rl} \, +\frac{Q_\rl}{1+2D}.
 \end{align}
 The ideal-gas pressure in the gaseous phase is balanced by a combination of an ideal-gas pressure and active stress in the liquid phase. As the liquid phase is denser, this requires that that the active stress will act in the opposite direction. For pusher cells, this is possible for $Q_{\l}<0$, {\it i.e.,} the cells are oriented in the $y$ direction, normal to the varying density profile. 
 
 The values of $\rho_\g$ and $\rho_\rl$ at co-existence define the binodal. We continue its derivation by inserting $\vecj$ in terms of $\tenQ_\rc$ in the equation for the nematic order to find
\begin{align}
    \label{eqs15}
  D\left(1+\frac{1}{4\left(1+2D\right)}\right)Q_\rc''=Q_\rc+\left(\sigma+\frac{Q_\rc}{1+2D}\right)\,g\left(Q_t\right), 
\end{align}
where the total nematic order is $Q_t=\beta_\rc Q_\rc+\beta_m Q_\m$. To simplify the equations, we further rescale lengths with $\sqrt{D\left(1+\frac{1}{4\left(1+2D\right)}\right)},$ such that the left-hand side of Eq.~(\ref{eqs15}) simplifies to $Q_\rc''$. Furthermore, we define the right-hand side of the equation as $F\left(\sigma,Q_\rc\right)$, such that
\begin{align}
    \label{eqs16}
  Q_\rc''=F\left(\sigma,Q_\rc\right).
  \end{align}

Equation~(\ref{eqs16}) has the same structure as Newton's equation, where the $x$ coordinate plays the role of time and $F$ plays the role of the force (see also~\cite{Solon2015}). The first integral (conservation of energy) yields 
\begin{align}
    \label{eqs17}
  E=\half Q_\rc '^2+U,
  \end{align}
where we have denoted the ``potential 
energy'' $U=-\int\D 
Q_\rc\,F\left(\sigma,Q_\rc\right).$ The binodal line describe 
the density and nematic order at 
macroscopically phase-separated 
states. Continuing the analogy to a 
Newtonian particle, this requires two 
$Q_\rc$ values that have the same 
energy (such that there is co-
existence) and where the force 
vanishes, $F=0$, such that it takes 
an infinite time to escape them. 
Without loss of generality, we take 
the energy value to be $E=0$. These 
two conditions set $Q_\rl$, the 
nematic order in the dense phase, as 
well as $-\sigma=\rho_{\g}$, the 
density in the isotropic gas phase. 
To summarize, we require that 
$Q_{\rc}=0, Q_{\l}$ are degenerate roots 
of $U$. 

\ra{Generally, co-existence between {\it finite-sized} domains with different $Q_\rc$ values is possible as long as they share the same value of $U$. The width of the domains is then determined by the values of $F$. }

\subsection{Cell-dominated and matrix-dominated limits}
It instructive to focus on two limiting cases: a cell-dominated case, where $\beta_\m=0$, and a matrix-dominated case, where $\beta_\rc=0$.

{\bf Cell-dominated case.} In the 
absence of a matrix, we denote 
$U\left(\beta_\m=0\right)=U_\rc$ and 
find that 
\begin{align}
    \label{eqs18aa}
    U_\rc \left(Q_\rc\right)&=\int _0 ^{Q_\rc} \D Q \left[\rho_\rc (Q) g\left(\beta_\rc Q\right)-Q\right].
\end{align}
Multiplying $U_\rc$ by $\beta_\rc ^2,$ 
we see that it is a function of 
$\beta_\rc \rho_\rc$ and $\beta_\rc 
Q_\rc$. This means that the phase 
diagram will be written in terms of 
lines of the form $\beta_\rc 
\rho_\rc=\mathrm{const.}$ 

The value of the nematic order at the binodal is found by 
differentiating $U_\rc$ and requiring $F=0$. Expanding for 
small $Q_\rc$ we find that $\beta_\rc 
Q_{\l}=-16/\left[3\left(1+2D\right)\beta_\rc\rho_\g 
\right]$. In particular, having found the spinodal at 
$\beta_\rc \rho_\ast=2$, we know that 
$\beta_{\rc}\rho_{\g}<2$. This means that 
$\beta_\rc Q_{\rl}<-8/\left[3\left(1+2D\right)\right]$. For small $D$ 
values, as we expect for living cells, we find that 
$\beta_\rc Q_{\l}<-2$. For such values, the linear 
expansion that we have used is not valid. It indicates 
that, in order to derive the phase diagram, $U$ should not 
be expanded around $Q_\rc=0$, but rather its full 
nonlinear form should be used and $Q_\rc$ shoud be found 
numerically.

The value of $\rho_\g=-\sigma$ at the binodal is found by requiring that $Q_\rc=0,Q_\rl$ have equal values of the ``potential energy'' $U_\rc$ with a vanishing ``force'' $F=0$. Then, the liquid branch of the binodal is found  by requiring that the stress is fixed, 
\begin{align}
    \label{eqs20a}
\beta_\rc\rho_{\l}=\beta_\rc\rho_\g+\frac{16}{3\left(1+2D\right)^2}\frac{1}{\beta_\rc\rho_\g}.
\end{align}
\\ {\bf Matrix-dominated case.} In the absence of cell-cell interactions, we denote $U\left(\beta_\rc=0\right)=U_\m$ and find that
\begin{align}
    \label{eqs21a}
    U_\m\left(Q_\rc\right)&=\int_0 ^{Q_\rc} \D Q \left[\rho_\rc (Q) g\left(\tilde{\beta_\m}q_\rc(Q)\right)-Q\right]\nonumber\\
    &\approx\int_0 ^{Q_\rc} \D Q \left[\left(-1+\frac{\tilde{\beta_\m}}{2}\right)Q-\frac{\tilde{\beta_\m}^3}{16\sigma^2}Q^3\right]\nonumber\\
    &=-\frac{\tilde{\beta_\m}^3}{64\sigma^2}Q_\rc^2\left(Q_\rc^2-16\sigma^2\frac{-2+\tilde{\beta_\m}}{\tilde{\beta_\m}^3}\right).
\end{align}
The potential is qualitatively different in this case. For $\tilde{\beta_\m}<2,$ it has a local maximum at $Q_\rc=0$ and no order forms. For $\tilde{\beta_\m}>2,$ however, $Q_\rc=0$ is a local minimum and there are two maxima at $Q_\rc=\pm2\sigma\sqrt{\left(-2+\tilde{\beta_\m}\right)\tilde{\beta_\m}^{-3}}$. Here we assumed that we are close to the transition $\tilde{\beta_\m}\approx2.$ 

The value of the intensive, nematic order parameter, $q_\rc=Q_\rc/\rho_\rc$ in this case, depends only on $\tilde{\beta_\m}$. This is obtained from the steady-state condition $q_\rc=g\left(\tilde{\beta_\m}q_\rc\right).$ The explicit dependence on $\tilde{\beta_\m}$ also infers that the phase diagram is not given by straight lines, as was in the cell-dominated case.

The potential $U_\m \left(\tilde{\beta_\m}\gtrapprox2\right)$ infers co-existence of nematic order along the $x$ and $y$ directions at any value of the density. This is possible because cells order at arbitrarily low densities. Then, cells aligned along the $x$ direction at very low densities can exert a positive active stress that matches $\sigma$.
Note that $\sigma$ here does not correspond to the density of the gas phase, but is simply the average cell density.  

\ra{The co-existence in this case is between finite-sized domains. To demonstrate this fact, we analytically calculate the energy difference between two solutions $q_\rc=\pm q$ of $F=0$ [Eq.~(4) of the main text]. We make use of the identity $Q_\rc=-\sigma q_\rc/\left[1+q/\left(1+2D\right)\right]$ to rewrite $U_\m$ as an integral over the intensive order parameter. We find that for $q>0,$ the energy difference $\Delta U_\m= U_\m\left(q\right)-U_\m\left(-q\right)$ is given by}
\ra{\begin{align}
   \label{eqs29a}
   \Delta U_\m&=\sigma^2\int_{-q}^{q}\D q'\left[-q'+g\left(\tilde{\beta_\m}q'\right)\right]\left[\left(1+\frac{q'}{1+2D}\right)^{-3}\right]\nonumber\\
   &=\sigma^2\int_0^{q}\D q'\left[q'-g\left(\tilde{\beta_\m}q'\right)\right]\left[\left(1-\frac{q'}{1+2D}\right)^{-3}-\left(1+\frac{q'}{1+2D}\right)^{-3}\right]\nonumber\\
   &<0.
\end{align}}
\ra{Here we have used the fact that $q'-g\left(\tilde{\beta_\m}q'\right)<0$ for $0<q'<q$. The above calculation means that finite-sized bands of $-q<q_\rc<0$ (in $y-$direction) can co-exist with positive $q_\rc>0$ (in $x-$direction).}

\ra{In the presence of a surface at $x=0$, the form of $U_\m$ allows for nematic order that is anchored along the $x$ direction to gradually change into its steady-state configuration along the $y$ direction in the bulk. This transition is characterized by a coherence length $l$ that, close to the ordering transition, diverges logarithmically with $\tilde{\beta_\m}-2$, as we show next. The length $l$ can also be regarded as the thickness of a pre-wetting layer. }

\ra{We consider a surface at $x=0$ that enforces a nematic order $q_\rc\left(x=0\right)=q>0$, where $F(q)=0$. Far away from the surface, the nematic order reaches the value $q_\rc=-q$, where the ``potential energy'' is at its global maximum, $U_\m(-q)=E$. We define $l$ as the distance over which the nematic order vanishes, $q\left(x=l\right)=0$. Note that the nematic order here is treated as a scalar. It does not rotate, but rather changes sign, inferring a disordered intermediate region.}

\ra{The thickness $l$ is given by}
\ra{\begin{align}
\label{eq29b}
l&=-\int_{Q_\rc'\left(Q_\rc=Q\right)} ^{Q_\rc'\left(Q_\rc=0\right)}\frac{\D Q_\rc'}{Q_\rc''}=-\int_{\sqrt{2\Delta U_\m}} ^{\sqrt{2E}}\frac{\D Q_\rc'}{F},    
\end{align}}
\ra{where $Q=\rho_\rc q$. Here we have used the equation (``Newton's law'') $Q_\rc''=F$ for the integrand  and the first integral (``conservation of energy'') $Q_\rc'^2/2+U_\m=E$ for the integration limits.}

\ra{We focus on $\tilde{\beta_\m}\gtrapprox2$ and expand $U_\m\approx-U_0 Q_\rc^2\left(Q_\rc^2-2Q^2\right),$ where $U_0>0$ is a prefactor. Conservation of energy then allows to relate between $Q_\rc$ and $Q_\rc'$ according to $Q_\rc'\approx\sqrt{2U_0}\left(Q_\rc^2-Q^2\right)$. We similarly expand $F\approx4U_0 Q_\rc\left(Q_\rc^2-Q^2\right)$ and find that, to leading order, $F\sim Q_\rc'$. Inserting back in Eq.~(\ref{eq29b}) yields }
\ra{\begin{align}
    \label{eq29b}
    l\sim\int_{\sqrt{2\Delta U_\m}} ^{\sqrt{2E}}\frac{\D Q_\rc'}{Q_\rc'}=\frac{1}{2}\ln\frac{E}{\Delta U_\m}.
\end{align}}

\ra{While $E/\Delta U_\m$ is always larger than unity, both $E$ and $\Delta U_\m$ vanish at $\tilde{\beta_\m}=2.$ For completeness, we find their scaling with $\tilde{\beta_\m}-2$ close to the transition. First, $E=U_\m\left(Q_\rc=Q\right)=U_0 Q^4\sim\left(\tilde{\beta_\m}-2\right)^2$. Second, expanding Eq.~(\ref{eqs29a}) yields $\Delta U_\m\sim\left(\tilde{\beta_\m}-2\right)q^3\sim\left(\tilde{\beta_\m}-2\right)^{5/2}.$ Overall,  we find that $l\sim-\ln\left(\tilde{\beta_\m}-2\right)$.}

\section{Non-linear angular dynamics}
\label{SMsec4}
Next, we focus on the angular dynamics of both the cells and matrix in the limits of large wavelengths. In the absence of a matrix, the cellular angle is a soft mode and its global rotation does not decay. The matrix, however, introduces a preferred axis, and any relative angle between the cells and matrix is expected to decay over time with a rate that is independent of system size. The relative angle closes via both cellular and matrix dynamics, and their interplay depends on the typical cellular and matrix rates.

We analyze the angular dynamics by writing the nematic tensors in terms of their modulus and angle
\begin{equation}
   \label{eqs29}
   \tenQ_\rc=\frac{Q_\rc}{2}\left(\begin{matrix}\cos2\phi_\rc&\sin2\phi_\rc\\\sin2\phi_\rc&-\cos2\phi_\rc\end{matrix}\right),\,\tenQ_\m=\frac{Q_\m}{2}\left(\begin{matrix}\cos2\phi_\m&\sin2\phi_\m\\\sin2\phi_\m&-\cos2\phi_\m\end{matrix}\right),\,\tenQ_t=\frac{Q_t}{2}\left(\begin{matrix}\cos2\phi_t&\sin2\phi_t\\\sin2\phi_t&-\cos2\phi_t\end{matrix}\right).
\end{equation}

The equations on $\tenQ_\rc$
and $\tenQ_\m$ can be thought of as equations on vectors that can be represented in polar coordinates. This results in equations on the moduli $Q_{\rc}$ and $Q_{\m}$ and the angles $\phi_\rc$ and $\phi_\m,$
respectively,
\begin{align}
   \label{eqs30}
\partial_{t}Q_{\rc} & =-Q_{\rc}+\rho_\rc g\left(Q_{t}\right)\cos2\left(\phi_\rc-\phi_t\right),\nonumber \\
2Q_{\rc}\partial_{t}\phi_\rc & =\rho_\rc g\left(Q_{t}\right)\sin2\left(\phi_\rc-\phi_t\right),\nonumber \\
\partial_{t}Q_{\m} & =-\rho_\rc k_{-}Q_{\m}+k_{+}Q_{\rc}\cos2\left(\phi_\rc-\phi_\m\right),\nonumber \\
2Q_{\m}\partial_{t}\phi_\m & =k_{+}Q_{\rc}\sin2\left(\phi_\rc-\phi_\m\right).
\end{align}
As the dynamics are invariant to global rotations of the systems,
the angle $\phi_t-\phi_\rc$ should depend only on $\phi_\rc-\phi_\m$. We
find it by taking $\phi_\rc=0$, such that $Q_{t}\sin2\phi_t=\beta_{\m}Q_{\m}\sin2\phi_\m$
and $Q_{t}\cos2\phi_t=\beta_{\rc}Q_{\rc}+\beta_{\m}Q_{\m}\cos2\phi_\m.$ This
yields
\begin{align}
   \label{eqs31}
\partial_{t}Q_{\rc} & =-Q_{\rc}+\frac{\rho_\rc}{Q_{t}}g\left(Q_{t}\right)\left[\beta_{\rc}Q_{\rc}+\beta_{\m}Q_{\m}\cos2\left(\phi_\rc-\phi_\m\right)\right],\nonumber \\
2Q_{\rc}\partial_{t}\phi_\rc & =-\frac{\rho_\rc}{Q_{t}}g\left(Q_{t}\right)\beta_{\m}Q_{\m}\sin2\left(\phi_\rc-\phi_\m\right),\nonumber \\
\partial_{t}Q_{\m} & =-\rho_\rc k_{-}Q_{\m}+k_{+}Q_{\rc}\cos2\left(\phi_\rc-\phi_\m\right),\nonumber \\
2Q_{\m}\partial_{t}\phi_\m & =k_{+}Q_{\rc}\sin2\left(\phi_\rc-\phi_\m\right).
\end{align}
In particular, we find that the relative angle decays according to
\begin{align}
   \label{eqs32}
2\partial_{t}\left(\phi_\rc-\phi_m\right) & =-\left[\beta_{\m}\rho_m\frac{q_{\m}}{q_{\rc}}\frac{g\left(Q_{t}\right)}{Q_{t}}+k_{+}\frac{\rho_\rc}{\rho_\m}\frac{q_{\rc}}{q_{\m}}\right]\sin2\left(\phi_\rc-\phi_m\right).
\end{align}
The first term on the right-hand side is the cellular contribution, while the second is the matrix contribution. \\
{\bf Analysis of the relaxation rates.} We focus on sufficiently ordered systems, such that we can neglect the dynamics in the densities and nematic order parameters, and focus only on angular dynamics. In this case, $q_\rc=g(Q_t)$, $q_\rc=q_\m$ and $\rho_\m=k_+/k_-$. Eq.~(\ref{eqs32}) then reduces to
\begin{align}
   \label{eqs33}
2\partial_{t}\left(\phi_\rc-\phi_m\right) & =-\left(\frac{\beta_{\m}\rho_{\m}}{\beta_{\m}\rho_{\m}+\beta_{\rc}\rho_{\rc}}+k_{-}\rho_{\rc}\right)\sin2\left(\phi_\rc-\phi_m\right).
\end{align}

The matrix rotation rate is  $k_- \rho_\rc$, because it is completely determined by the degradation rate (recall that we have omitted the $k_0$ terms from our analysis), while the cellular rate is of order $k \beta_\m  \rho_\m/\left(\beta_\m  \rho_\m+\beta_\rc  \rho_\rc\right)$, where we have reintroduced the timescale $1/k$. 

\ra{{\bf Arrested domain coarsening.} As is discussed in the main text, cellular alignment with the matrix is expected to arrest the coarsening of differently-oriented cellular domains. Explicitly, we expect domains of typical size $l$ to remain frozen as long as $k \beta_\m  \rho_\m/\left(\beta_\m  \rho_\m+\beta_\rc  \rho_\rc\right)\gg D_t/l^2,\,k_-\rho_\rc$. Here $D_t$ is the total translational diffusion coefficient. In our model, it is given by $D_t=D+v^2/(4k)$. }

\begin{figure*}[ht]
\centering
\includegraphics[width=0.95\textwidth]{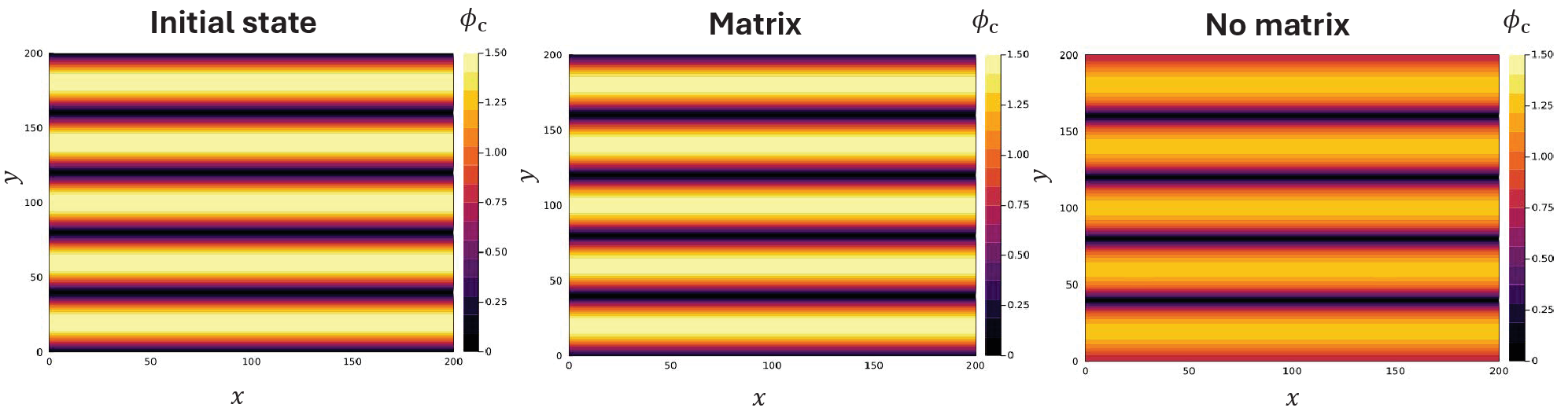}
\caption{\ra{(Color online) Cellular angles $\phi_\rc$ obtained from numerical solutions of Eq. (2) of the main text, $10^4$ time steps after common initial state (left panel) in the form of alternating bands, according to Eq.~(\ref{eqs34}). We compare between a matrix-dominated case ($\beta_\rc=0,\beta_\m=10,$ center panel) and cell-dominated case ($\beta_\rc=10,\beta_\m=0,$ right panel). The values used are: $L=200$ (system size),  $D=0.5,$ $\rho_\rc^0=1$ (initial cellular density), $k_+=1,$ and $k_-=0.1.$. Periodic boundary conditions are used. The matrix-dominated snapshot is barely distinguishable from the initial state, while the cell-dominated snapshot displays  coarsening.}}
\label{figs1}
\end{figure*}

\ra{We test this prediction by numerically solving the hydrodynamic equations [Eq.~(2) of the main text]. We consider initial conditions of fully aligned cells and matrix ($q_\rc=q_\m=1$), whose direction alternates along the $y$-direction, according to }
\begin{align}
\label{eqs34}
    \phi_\rc=\phi_\m&=\frac{\pi}{2}\cos\frac{2\pi}{l}y.
\end{align}
\ra{Here, $l$ is the band width. We integrate the equations for $10^4$ time steps in the two limits of cell-dominated interaction ($\beta_\m=0$) and matrix dominated interaction ($\beta_\rc=0$). In accordance with our predictions, the domains coarsen in the cell-dominated case and display negligible dynamics in the matrix-dominated case, as is evident from Fig.~\ref{figs1}. }

\end{document}